\documentclass[a4paper,oneside]{article}
\usepackage[english]{babel}
\usepackage{amsmath}
\usepackage{graphicx}
\usepackage[colorlinks=true, allcolors=blue]{hyperref}
\usepackage[utf8]{inputenc}
\usepackage{amssymb}
\usepackage{amsfonts}
\usepackage[warn]{mathtext}
\usepackage{float}
\usepackage{layout}
\usepackage{revsymb}
\usepackage{multirow}
\usepackage{authblk}
\title{The generation of high-energy electron-positron pairs during the Breit-Wheeler resonant process in a strong field of an X-ray electromagnetic wave}
\author[1]{S.P. Roshchupkin}
\author[1]{V.D. Serov}
\author[1]{V.V. Dubov}
\date{}
\affil[1]{Peter the Great St. Petersburg Polytechnic University,
195251, St-Petersburg, Russian Federation, Russia}
\begin{document}

\maketitle
\begin{abstract}
    The Breit-Wheeler resonant process has been theoretically studied in a strong X-ray electromagnetic wave field under conditions when the energy of one of the initial high-energy gamma quanta passes into the energy of a positron or electron. These resonant conditions have been studied in detail. Analytical formulas for the resonant differential cross section of Channels A and B of the reaction are obtained. It is shown that the resonant differential cross section significantly depends on the value of the characteristic Breit-Wheeler energy, which is determined by the parameters of the electromagnetic wave and the initial gamma quanta. With a decrease in the characteristic Breit-Wheeler energy, the resonant cross section increases sharply and may exceed the corresponding non-resonant cross section by several orders of magnitude.
\end{abstract}
\section{Introduction}
\par Obtaining narrow beams of ultrarelativistic positrons (electrons) is an important scientific problem. The Breit-Wheeler process (BWP) is well known, which involves the production of electron-positron pairs through the interaction of two gamma quanta \cite{1}. In this process, the energy spectrum of the generated electrons and positrons is continuous (the energy of the initial gamma quanta is distributed between the electron and the positron). With the advent of high-intensity lasers \cite{2}-\cite{11}, the study of quantum electrodynamics (QED) processes in external electromagnetic fields has been widely explored (see, for example, reviews \cite{12}-\cite{19}, monographs \cite{20}-\cite{22}, and articles \cite{23}-\cite{57}). Moreover, these processes can occur both resonantly and non-resonantly. Resonant behavior of high-order QED processes are associated with the entering of an intermediate particle onto the mass shell. These resonances were first investigated by Oleinik \cite{23,24}. As a result, higher-order processes by the fine-structure constant effectively reduce into several consecutive lower-order processes (see reviews \cite{5,13,14}, monographs \cite{20,21,22}), as well as recent articles \cite{34}-\cite{39}, \cite{52}). It is important to note that the probability of resonant processes can significantly exceed the corresponding probabilities of non-resonant processes.
\par Currently, there is a considerable body of work dedicated to studying BWP in an external electromagnetic field (see, for example, \cite{40}-\cite{52}). It is important to distinguish between external field-stimulated BWP (a first-order process by the fine-structure constant) and external field-assisted BWP (a second-order process by the fine-structure constant).
\begin{figure}[H]
    \centering
\begin{minipage}{.49\textwidth}
    \centering
   \includegraphics[width=0.8\linewidth]{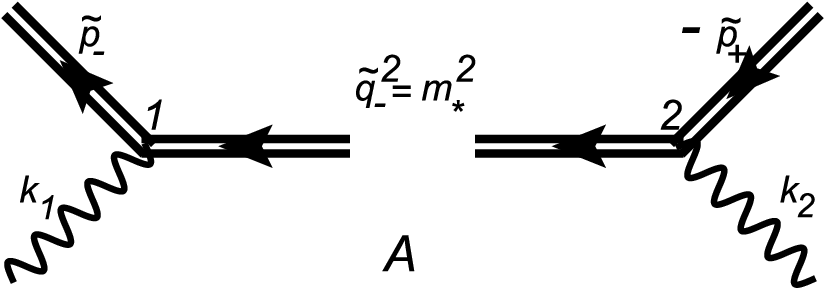}
\end{minipage}
\hfill
\begin{minipage}{.49\textwidth}
    \centering
   \includegraphics[width=0.8\linewidth]{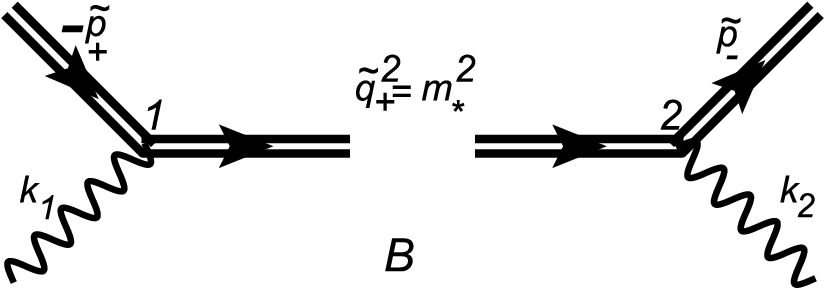}
\end{minipage}
\caption{Feynman diagrams of the resonant electron–positron-pair production by two gamma quanta in an external
field, Channels A and B; for Channels A' and B' $k_1\leftrightarrow k_2$.}
\label{fig1}
\end{figure}
\par In the recent article \cite{52}, the resonant external field-assisted BWP in a strong field of a monochromatic circularly-polarized wave, propagating along the $z$ axis, was studied. It is important to note that under the conditions of resonance and absence of interference from different reaction channels, the original second-order process effectively reduces into two first-order processes: stimulated external field BWP and stimulated external field Compton effect (CE) \cite{12} (see Fig.\ref{fig1}). The resonant BWP for high-energy initial gamma quanta and ultrarelativistic final electron-positron pairs was considered
\begin{equation}\label{1}
\hbar\omega_{1,2}\gg mc^2,\quad E_{\pm}\gg mc^2.
\end{equation}
Here $\hbar\omega_{1,2}$ and $E_{\pm}$ are the energies of the initial gamma quanta and the final positron or electron. At the same time, all particles (the initial gamma quanta and the final electron-positron pair) move in a narrow cone away from the direction of wave propagation
\begin{equation}\label{2}
\theta_{j\pm}\equiv\angle(\mathbf{k}_j, \mathbf{p}_{\pm})\ll1,\quad\theta_{i}\equiv\angle(\mathbf{k}_1, \mathbf{k}_2)\ll1,
\end{equation}
\begin{equation}\label{3}
\theta_j\equiv\angle(\mathbf{k}_j, \mathbf{k})\sim1,\quad j=1,2;\quad\theta_{\pm}\equiv\angle(\mathbf{p}_{\pm},\mathbf{k})\sim1,\quad\theta_{1,2}\approx\theta_{\pm}\equiv\theta.
\end{equation}
\par Here, $\mathbf{p}_{\pm}$ are momenta of the positron or the electron. It was assumed that the classical relativistic-invariant parameter satisfies the following condition \cite{12}, \cite{35}-\cite{39}:
\begin{equation}\label{4}
\eta=\frac{eF\lambdabar}{mc^2}\ll\eta_{max},\quad \eta_{max}=\mathrm{min}\left(\frac{E_{\pm}}{mc^2}\right).
\end{equation}
Here $e$ and $m$ are the charge and mass of the electron, $F$ and $\lambdabar=c/\omega$ are the electric field strength and wavelength, $\omega$ is the frequency of the wave.
\par In the article \cite{52}, it was shown that the resonant energy of positron and electron has a discrete spectrum. The outgoing angles of the electron and positron are uniquely related and determine their resonant energies. In addition, the resonant energies of the electron-positron pair for each reaction channel are determined by the quantum parameters $\varepsilon_{1C(r')}$ (at the first vertex for the external field-stimulated CE) and $\varepsilon_{2BW(r)}$ (at the second vertex for the external field-stimulated BWP):
\begin{equation}\label{5}
\varepsilon_{1C(r')}=r'\varepsilon_{1C},\quad\varepsilon_{1C}=\frac{\hbar\omega_1}{\hbar\omega_{C}},\quad\hbar\omega_{C}=\frac{1}{4}\hbar\omega_{BW},
\end{equation}
\begin{equation}\label{6}
\varepsilon_{2BW(r)}=r\varepsilon_{2BW}\ge1,\quad\varepsilon_{2BW}=\frac{\hbar\omega_2}{\hbar\omega_{BW}},
\end{equation}
\begin{equation}\label{7}
\hbar\omega_{BW}=\frac{(mc^2)^2(1+\eta^2)}{\hbar\omega\sin^2(\theta/2)}.
\end{equation}
In expression (\ref{5}), $r'=1,2,3\dots$ represents the number of emitted photons of the wave at the first vertex, while $\hbar\omega_{C}$ is the characteristic energy of the external field-stimulated CE. In equation (\ref{6}), $r$ represents the number of photons absorbed at the second vertex, while $\hbar\omega_{BW}$ (\ref{7}) is the characteristic energy of the external field-stimulated BWP. Unlike the first vertex, the number of absorbed photons from the wave $r$ at the second vertex significantly depends on the relationship between the energy of the second gamma quantum and the characteristic energy $\hbar\omega_{BW}$. Thus, if the quantum parameter is $\varepsilon_{2BW}<1\quad(\omega_2<\omega_{BW})$, then the number of absorbed photons of the wave at the second vertex starts from a certain minimum value: $r\ge r_{min}=\lceil\varepsilon_{2BW}^{-1}\rceil>1$ (see condition (\ref{6}) for parameter $\varepsilon_{2BW(r)}$). However, if the quantum parameter is $\varepsilon_{2BW}\ge1\quad(\omega_2\ge\omega_{BW})$, then the number of absorbed photons starts from one: $r=1,2,3\dots$.
\par It should be emphasized that the article \cite{52} thoroughly investigated the case where the following conditions were imposed on the energies of the initial gamma quanta within framework (\ref{1}):
\begin{equation}\label{8}
\omega_2\sim\omega_{BW},\omega_1\ll\omega_{BW},\omega_1\sim\omega_C\qquad(\varepsilon_{2BW}\sim1,\varepsilon_{1BW}\ll1,\varepsilon_{1C}\sim1).
\end{equation}
It should be noted that under these conditions, the exchange resonant reaction channels $(k_1\longleftrightarrow k_2)$ were suppressed and not considered.
\par It is important to highlight that from a physical point of view, the most interesting case, which was not studied in article \cite{52}, is when the energy of the second gamma quantum significantly exceeds the characteristic energy of the external field-stimulated BWP, while the energy of the first gamma quantum is on the order of or much smaller than the characteristic energy CE. In this case, instead of relationships (\ref{8}), the following conditions are obtained
\begin{equation}\label{9}
\omega_2\gg\omega_{BW},\omega_1\ll\omega_{BW},\omega_1\sim\omega_C\qquad(\varepsilon_{2BW}\gg1,\varepsilon_{1BW}\ll1,\varepsilon_{1C}\sim1);
\end{equation}
\begin{equation}\label{10}
\omega_2\gg\omega_{BW},\omega_1\ll\omega_C\qquad(\varepsilon_{2BW}\gg1,\varepsilon_{1C}\ll1).
\end{equation}
\par It is important to note that the characteristic Breit-Wheeler energy  $\hbar\omega_{BW}$ (\ref{7}), at a fixed parameter $\eta$ value, is inversely proportional to the energy of the photon of the electromagnetic wave. As the photon energy of the wave increases from optical frequencies to X-ray frequencies, the characteristic Breit-Wheeler energy decreases while the intensity of the wave increases. Let us estimate the characteristic energy of the external field-stimulated BWP $\hbar\omega_{BW}$ (\ref{7}). In deriving the estimate, we will consider frequencies of the electromagnetic wave in the X-ray range, as well as the angle $\theta=\pi$:
\begin{equation}\label{11}
   \hbar\omega_{BW}=\left\{ \begin{array}{rcl}
17.4\,\mbox{GeV}&\mbox{if}&\omega=30\,\mbox{eV},I=1.675\cdot10^{21}\,\mbox{Wcm}^{-2}\,(\eta^2=1)\\
522\,\mbox{MeV}&\mbox{if}&\omega=1\,\mbox{keV},I=1.861\cdot10^{24}\,\mbox{Wcm}^{-2}\,(\eta^2=1)\\
522\,\mbox{MeV}&\mbox{if}&\omega=10\mbox{keV},I=3.536\cdot10^{27}\,\mbox{Wcm}^{-2}\,(\eta^2=19)
   \end{array}\right.
\end{equation}
\par We note that in equation (\ref{11}), for the frequencies of the X-ray wave $\omega=30$\,eV and $\omega=1$\,keV the wave parameters were chosen such that the characteristic energy of the BWP decreases with increasing wave intensity, from $\hbar\omega_{BW}=17.4$\,GeV to $\hbar\omega_{BW}=522$\,MeV. At the same time, for X-ray frequencies ranging from $\omega=1$\,keV to $\omega=10$\,keV, the wave parameters were chosen such that the characteristic energy of the BWP $\hbar\omega_{BW}=522$\,MeV remains unchanged with increasing wave intensity.
\par In this article, we will study the resonant BWP in a strong X-ray field under the conditions for the initial gamma quanta energies specified by equations (\ref{9}) and (\ref{10}). As will be shown later, under these conditions, the resonant energy of the positron (for Channel A) or electron (for Channel B) tends to be the energy of the high-energy second gamma quantum with the greatest probability.
\par Subsequently, the relativistic system of units is used: $c=\hbar=1$.
\section{Resonant energies of the positron (electron)}
\par Oleinik resonances occur when an intermediate electron or positron in an external field enters the mass shell. As a result, for Channels A and B, we obtain (see Fig.\ref{fig1}):
\begin{equation}\label{12}
\widetilde{q}_-^2=m_*^2,\quad\widetilde{q}_-=k_2-\widetilde{p}_++rk=\widetilde{p}_--k_1+r'k,
\end{equation}
\begin{equation}\label{13}
\widetilde{q}_+^2=m_*^2,\quad\widetilde{q}_+=k_2-\widetilde{p}_-+rk=\widetilde{p}_+-k_1+r'k.
\end{equation}
In expressions (\ref{12})-(\ref{13}), $k_{1,2}=(\omega_{1,2},\mathbf{k_{1,2}})$ and $k=(\omega,\mathbf{k})$ are 4-momenta of the first and second gamma quanta and the external wave photon, $\widetilde{p}_{\pm}=(\widetilde{E}_{\pm}, \mathbf{\widetilde{p}_{\pm}})$ and $\widetilde{q}_{\mp}$ are 4-quasimomenta of the final positron (electron) and intermediate electron (positron), $m_*$ is the effective mass of the electron in the
field of a circularly polarized wave \cite{12}:
\begin{equation}\label{14}
\widetilde{p}_{\pm}=p_{\pm}+\eta^2\frac{m^2}{2(kp_{\pm})}k,\quad\widetilde{q}_{\mp}=q_{\mp}+\eta^2\frac{m^2}{2(kq_{\mp})}k,
\end{equation}
\begin{equation}\label{15}
\widetilde{p}_{\pm}^2=m_*^2,\quad m_*=m\sqrt{1+\eta^2},
\end{equation}
\begin{equation}\label{16}
\widetilde{p}_{i,f}^2=m_*^2,\quad m_*=m\sqrt{1+\eta^2}.
\end{equation}
In expression (\ref{14}), $p_{\pm}=(E_{\pm}, \mathbf{p_{\pm}})$ --- 4-momenta of the positron (electron).
\par In this article, within the framework of conditions (\ref{9}) and (\ref{10}), we will consider the energies of the second gamma quantum $\omega_2\lesssim10^3$ GeV , as well as the range of X-ray frequencies $10\,\mbox{eV}\lesssim\omega\lesssim10^5\,\mbox{eV}$. At the same time, we will consider the intensities of the electromagnetic wave to be significantly smaller than the critical intensities of the Schwinger \cite{55,19} ($I\ll I_*\sim10^{29}$ W/cm$^2$).
\par Considering equations (\ref{1})-(\ref{4}), (\ref{12}), (\ref{13}), it is straightforward to obtain the resonant energies of positron $E_{+(r)}$ (for Channel A) or electron $E_{-(r)}$ (for Channel B) at the second vertex, where the external field-stimulated BWP occurs (see Fig.\ref{fig1}):
\begin{equation}\label{17}
    E_{\pm(r)}=\frac{\omega_2}{2(\varepsilon_{2BW(r)}+\delta^2_{2\pm})}\left[\varepsilon_{2BW(r)}+\sqrt{\varepsilon_{2BW(r)}(\varepsilon_{2BW(r)}-1)-\delta^2_{2\pm}}\right],
\end{equation}
where the ultrarelativistic parameter
\begin{equation}\label{18}
\delta_{2\pm}^2=\frac{\omega_2^2}{(2m_*)^2}\theta_{2\pm}^2,
\end{equation}
associated with the outgoing angle of the positron (electron) relative to the momentum of the second gamma quantum, can vary in the interval:
\begin{equation}\label{19}
0\le\delta^2_{2\pm}\le\delta^2_{2max},\quad\delta^2_{2max}=\varepsilon_{2BW(r)}(\varepsilon_{2BW(r)}-1).
\end{equation}
It should be noted that the quantity (\ref{17}) determines the maximum energy of the positron (electron). There is also a solution with a minus sign in front of the square root, which determines the minimum value of the resonant energy. However, the minimum resonant energy is unlikely \cite{52}. Under the conditions (\ref{9}), (\ref{10}), the quantum parameter $\varepsilon_{2BW(r)}=r\varepsilon_{2BW}\gg1\,(r\ge1)$. As a result, the resonant energy of the positron or electron (\ref{17}) will be close to the energy of the high-energy second gamma quantum:
\begin{equation}\label{20}
E_{\pm(r)}\approx\omega_2\left[1-\frac{(1+4\delta^2_{2\pm})}{4r\varepsilon_{2BW}}\right]\longrightarrow\omega_2\quad(\delta^2_{2\pm}\ll r\varepsilon_{2BW}).
\end{equation}
\begin{figure}[H]
\begin{center}
\center{\includegraphics[width=0.5\linewidth]{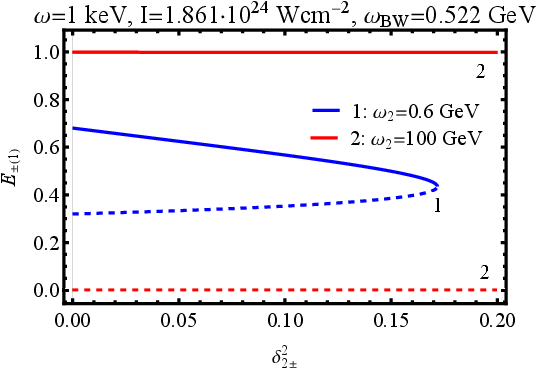}} 
\caption{The dependence of the resonant energy of the positron (electron) (in units of $\omega_2$) on the ultrarelativistic parameter $\delta^2_{2\pm}$  for a single absorbed photon of the wave with significantly different energies of the second gamma quantum.}
\label{fig2}
\end{center}
\end{figure}
\par Figure \ref{fig2} illustrates the resonant energy of the positron (for Channel A) or electron (for Channel B) as a function of the corresponding ultrarelativistic parameter $\delta^2_{2\pm}$ for the energies of the second gamma quantum $\omega_2=600$\,MeV and $\omega_2=100$\,GeV. This graph presents two solutions for the resonant energy, corresponding to the "plus" (solid curves) and "minus" (dashed curves) signs before the square root in equation (\ref{17}). As mentioned earlier, only the maximum energy of the positron (electron) will be used in the future (solid curves in Fig.\ref{fig2}). From Figure \ref{fig2}, it can be observed that there is a significant difference in the resonant energies of the positron (electron) for these second gamma quantum energies. Unlike the case with $\omega_2=600$\,MeV, the resonant energy of the positron (electron) for the case with $\omega_2=100$\,GeV tends to converge to the energy of the second gamma quantum at small values of the parameter $\delta^2_{2\pm}$ (see solid curve 2 in Fig.\ref{fig2}, as well as formula (\ref{20})).
\par Considering equations (\ref{1})-(\ref{4}), (\ref{12}), (\ref{13}), it is straightforward to obtain the resonant energies of electron $E_{-(r')}$ (for Channel A) or  positron $E_{+(r')}$ (for Channel B) at the first vertex, where the external field-stimulated CE occurs (see Fig.\ref{fig1}):
\begin{equation}\label{21}
     E_{\mp(r')}=\frac{\omega_1}{2(\varepsilon_{1C(r')}-\delta^2_{1\mp})}\left[\varepsilon_{1C(r')}+\sqrt{\varepsilon_{1C(r')}^2+4(\varepsilon_{1C(r')}-\delta^2_{1\mp})}\right].
\end{equation}
Here  the ultrarelativistic parameter, associated with the outgoing angle of the electron (positron) relative to the momentum of the first gamma quantum, is defined:
\begin{equation}\label{22}
\delta_{1\mp}^2=\frac{\omega_1^2}{m_*^2}\theta_{1\mp}^2.
\end{equation}
It should be noted that there are no constraints on the parameter $\varepsilon_{1C(r')}$ for the external field-stimulated CE. Therefore, this process occurs for any number of emitted photons of the wave $r'\ge1$. At the same time, $\delta_{1\mp}^2<\varepsilon_{1C(r')}$.
\par It is important to note that in the conditions of resonance (\ref{12}), (\ref{13}) for each of the reaction channels, the resonant energies of the positron and electron are determined by different physical processes: external field-stimulated BWP (\ref{17}) and CE (\ref{21}). At the
same time, the energies of the electron-positron pair are related to each other by the general law of
conservation of energy
\begin{equation}\label{23}
    E_++E_-\approx\omega_1+\omega_2.
\end{equation}
It should be noted that in equation (\ref{23}) we have neglected a small correction term $\sim|r-r'|\omega/\omega_2\ll1$. Taking into account relations (\ref{20}) and (\ref{21}), as well as the law of energy conservation (\ref{23}) for Channels A and B, we obtain the following equation relating the outgoing angles of the positron and electron within the conditions (\ref{9}), (\ref{10}):
\begin{equation}\label{24}
\delta^2_{1\mp}=\varepsilon_{1C}\left\{r'-\frac{r\varepsilon_{1C}}{1+4\delta^2_{2\pm}+r\varepsilon_{1C}}\left[r'+\frac{r}{1+4\delta^2_{2\pm}+r\varepsilon_{1C}}\right]\right\}.
\end{equation}
Here, the upper (lower) sign corresponds to Channel A (B). In relation (\ref{24}), to the left of the equality sign is the ultrarelativistic parameter responsible for the outgoing angle of electron (positron) relative to the momentum of the first gamma quantum, and to the right of the equality sign is the function of the ultrarelativistic parameter $\delta^2_{2\pm}$, which corresponds to the outgoing angle of positron (electron) relative to the momentum of the second gamma quantum. Under the given parameters $\varepsilon_{1C}$ and $(r,r')$, the relation (\ref{24}) uniquely determines the outgoing angles of electron and positron, and therefore their resonant energies. If we substitute expression (\ref{24}) into relation (\ref{21}), we obtain that under conditions (\ref{9}), the energy of the electron (for Channel A) or positron (for Channel B) is determined by the expression (also see relations (\ref{20}), (\ref{23})):
\begin{equation}\label{25}
E_{\mp}\approx\omega_1+\frac{1+4\delta^2_{2\pm}}{r}\omega_C.
\end{equation}
Moreover, under conditions (\ref{10}), the expression (\ref{25}) simplifies:
\begin{equation}\label{26}
E_{\mp}\approx\frac{1+4\delta^2_{2\pm}}{r}\omega_C.
\end{equation}
\par Thus, the resonant energy of the electron-positron pair in conditions (\ref{9}) is determined by equations (\ref{20}), (\ref{25}), and in conditions (\ref{10}) by equations (\ref{20}), (\ref{26}). It is important to emphasize that in conditions (\ref{9}), (\ref{10}), the energy of the positron (for Channel A) or electron (for Channel B) is determined by the energy of the second high-energy gamma quantum, while the energy of the electron (for Channel A) or positron (for Channel B) in conditions (\ref{10}) is primarily determined not by the energy of the first gamma quantum, but by the characteristic energy of CE.

\section{Maximum Breit-Wheeler resonant cross-section}
\par In the conditions (\ref{9}) and (\ref{10}), in Channels A and B the external field-stimulated BWP proceeds with the number of absorbed photons of the wave $r\ge1$, and for exchange resonant diagrams A' and B' the number of absorbed photons is $r\ge r_{1min}=\lceil\varepsilon_{1BW}^{-1}\rceil\gg1$. Therefore, the resonant Channels A' and B' will be suppressed, and it is sufficient to consider only two resonant Channels, A and B. It should also be noted that for Channel A, the resonant energy of the electron-positron pair is determined by the outgoing angle of the positron relative to the momentum of the second gamma quantum, while for Channel B --- by the outgoing angle of the electron relative to the momentum of the second gamma quantum. Therefore, Channels A and B are distinguishable and do not interfere with each other.
\par In the article \cite{52}, a general relativistic expression for the resonant BWP modified by a strong electromagnetic field was obtained. The resonance infinity, which occurs in the field of a plane monochromatic wave, was eliminated by the Breit-Wigner procedure \cite{57}, \cite{36}-\cite{39}. As a result, the maximum resonant differential cross section (at the point of maximum Breit-Wigner distribution) for Channel A (with a "plus" sign) and Channel B (with a "minus" sign) has the following form:
\begin{equation}\label{27}
R_{\pm(rr')}=\frac{d\sigma_{rr'}^{max}}{d\delta^2_{2\pm}}=r^2_ec_{\eta i}\Psi_{\pm(rr')}.
\end{equation}
Here, $r_e=e^2/m$ is the classical electron radius, and the function $c_{\eta i}$ is determined by the initial setup parameters:
\begin{equation}\label{28}
c_{\eta i}=\frac{2(4\pi)^3}{\alpha^2K^2(\varepsilon_{1C})}\left(\frac{m}{\delta_{\eta i}\omega_{BW}}\right)^2\approx0.745\cdot10^8\left(\frac{m}{\delta_{\eta i}\omega_{BW}K(\varepsilon_{1C})}\right)^2,
\end{equation}
where $\alpha$ is the fine-structure constant, and $\delta^2_{\eta i}$ is the ultrarelativistic parameter, which determines the angle between the momenta of the initial gamma quanta (see equation (\ref{2}));
\begin{equation}\label{29}
    \delta^2_{\eta i}=\frac{\omega_1\omega_2}{m^2_*}\theta^2_{i}.
\end{equation}
The function $K(\varepsilon_{1C})$ is defined by the total probability (per unit of time) of the external field-stimulated CE on
the intermediate electron (for Channel A) or positron (for Channel B):
\begin{equation}\label{30}
K(\varepsilon_{1C})=\sum_{s=1}^{\infty}\int_0^{s\varepsilon_{1C}}\frac{du}{(1+u)^2}K(u,s\varepsilon_{1C}).
\end{equation}
Here, the function $K(u,s\varepsilon_{1C})$ is determined by the expressions:
\begin{equation}\label{31}
    K(u,s\varepsilon_{1C})=-4J^2_{s}(y_{1(s)})+\eta^2\left[2+\frac{u^2}{1+u}\right](J^2_{s-1}+J^2_{s+1}-2J^2_{s}),
\end{equation}
\begin{equation}\label{32}
    y_{1(s)}=2s\frac{\eta}{\sqrt{1+\eta^2}}\sqrt{\frac{u}{s\varepsilon_{1C}}\left(1-\frac{u}{s\varepsilon_{1C}}\right)}.
\end{equation}
In equation (\ref{27}), functions $\Psi_{\pm(rr')}$ determine the spectral-angular distribution of the generated electron-positron pair:
\begin{equation}\label{33}
\Psi_{\pm(rr')}=\frac{x_{\pm(r)}}{\varepsilon^2_{2BW}(1-x_{\pm(r)})}K_{1\mp(r')}P_{2\pm(r)}.
\end{equation}
In expression (\ref{33}), the function $P_{2\pm(r)}$ determines the probability of the external field-stimulated BWP, and the function $K_{1\mp(r')}$ determines the probability of the external field-stimulated CE \cite{12}:
\begin{equation}\label{34}
    P_{2\pm(r)}=J^2_{r}(\gamma_{2\pm(r)})+\eta^2(2u_{2\pm(r)}-1)\left[\left(\frac{r^2}{\gamma_{2\pm(r)}^2}-1\right)J^2_{r}+J'^2_{r}\right],
\end{equation}
\begin{equation}\label{35}
    K_{1\mp(r')}=-4J^2_{r'}(\gamma_{1\mp(r')})+\eta^2\left[2+\frac{u_{1\mp(r')}^2}{1+u_{1\mp(r')}}\right](J^2_{r'-1}+J^2_{r'+1}-2J^2_{r'}).
\end{equation}
The arguments of the Bessel functions have the following form:
\begin{equation}\label{36}
    \gamma_{2\pm(r)}=2r\frac{\eta}{\sqrt{1+\eta^2}}\sqrt{\frac{u_{2\pm(r)}}{v_{2\pm(r)}}\left(1-\frac{u_{2\pm(r)}}{v_{2\pm(r)}}\right)},
\end{equation}
\begin{equation}\label{37}
    \gamma_{1\mp(r')}=2r'\frac{\eta}{\sqrt{1+\eta^2}}\sqrt{\frac{u_{1\mp(r')}}{v_{1\mp(r')}}\left(1-\frac{u_{1\mp(r')}}{v_{1\mp(r')}}\right)}.
\end{equation}
Here, the relativistic-invariant parameters are equal to:
\begin{equation}\label{38}
    u_{1\mp(r')}=\frac{(k_1k)}{(p_{\mp}k)}\approx\frac{\omega_1}{E_{\mp(r')}}\approx\frac{\omega_1}{\omega_1+\omega_2-E_{\pm(r)}},
\end{equation}
\begin{equation}\label{39}
  v_{1\mp(r')}=\frac{2r'(q_{\mp}k)}{m^2_*}\approx\varepsilon_{1C(r')}\left(\frac{E_{\mp(r')}}{\omega_1}-1\right)\approx\varepsilon_{1C(r')}\frac{\omega_2-E_{\pm(r)}}{\omega_1},
\end{equation}
\begin{equation}\label{40}
    u_{2\pm(r)}=\frac{(k_2k)^2}{4(p_{\pm}k)(q_{\mp}k)}\approx\frac{\omega_2^2}{4E_{\pm(r)}\left(\omega_2-E_{\pm(r)}\right)},\quad v_{2\pm(r)}=r\frac{(k_2k)}{2m_*^2}\approx\varepsilon_{2BW(r)}.
\end{equation}
It should be noted that in the right-hand sides of equations (\ref{38}) and (\ref{39}), the general law of conservation of energy (\ref{23}) has been taken into account. Therefore, the relativistic-invariant parameters for the external field-stimulated CE $u_{1\mp(r')}$ and $v_{1\mp(r')}$ can be expressed in terms of the energy of the positron (electron) for the external field-stimulated BWP.
\par
\begin{figure}[H]
\begin{minipage}[h]{0.325\linewidth}
\center{\includegraphics[width=0.99\linewidth]{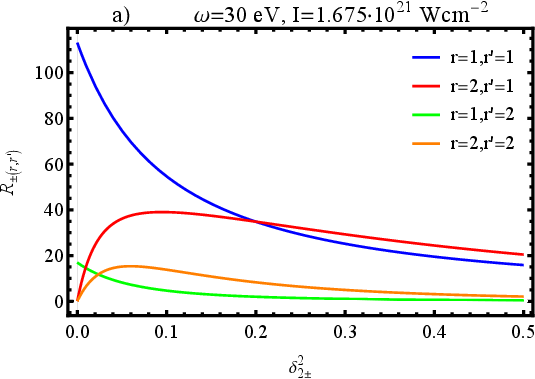}}
\end{minipage}
\hfill
\begin{minipage}[h]{0.325\linewidth}
\center{\includegraphics[width=0.99\linewidth]{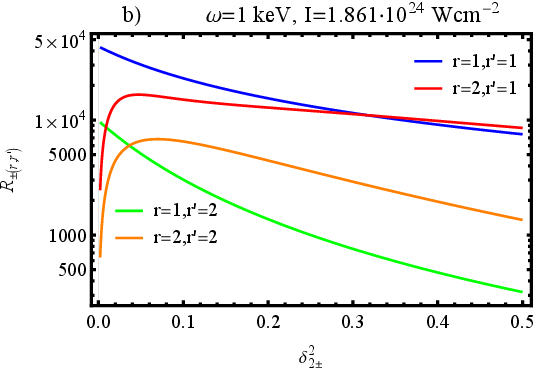}}
\end{minipage}
\hfill
\begin{minipage}[h]{0.325\linewidth}
\center{\includegraphics[width=0.99\linewidth]{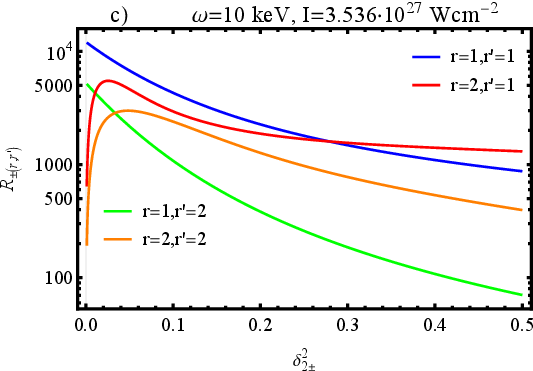}}
\end{minipage}
\caption{The dependence of the maximum resonant differential cross section (\ref{27}), (\ref{28}), (\ref{33}) (in units of $r_e^2$) on the square of the positron outgoing angle (Channel A) or the square of the electron outgoing angle (Channel B) for various intensities of the X-ray wave and the
characteristic energy of BWP, as well as the numbers of absorbed ($r$) and emitted ($r'$) photons in conditions (\ref{8}) for the energies of the initial gamma quanta. The corresponding energies:  Fig.3a) -- $\omega_{BW}=17.4\,\mbox{GeV},\,\omega_2=35\mbox{GeV},\omega_1=1\,\mbox{GeV}$; Fig.3b) and 3c) -- $\omega_{BW}=522\,\mbox{MeV},\,\omega_2=1\mbox{GeV},\omega_1=50\,\mbox{MeV}$.}
\label{fig3}
\end{figure}

\begin{figure}[H]
\begin{minipage}[h]{0.325\linewidth}
\center{\includegraphics[width=0.99\linewidth]{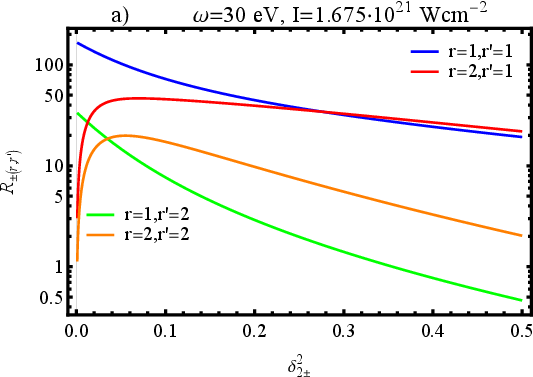}}
\end{minipage}
\hfill
\begin{minipage}[h]{0.325\linewidth}
\center{\includegraphics[width=0.99\linewidth]{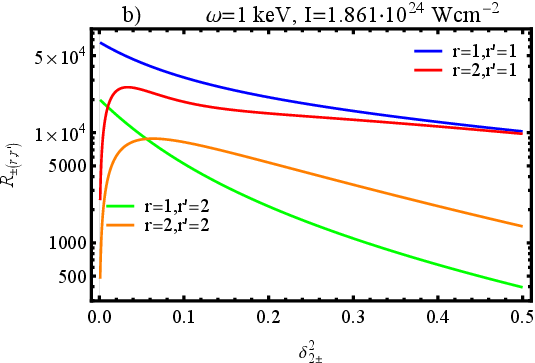}}
\end{minipage}
\hfill
\begin{minipage}[h]{0.325\linewidth}
\center{\includegraphics[width=0.99\linewidth]{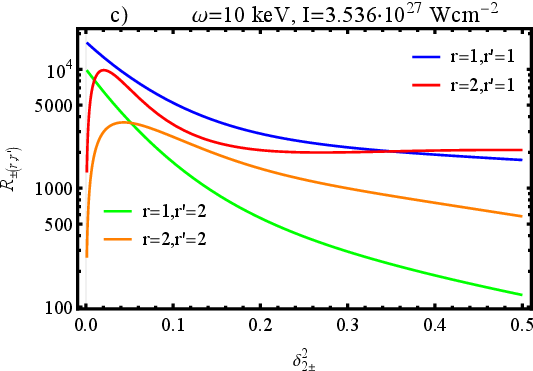}}
\end{minipage}
\caption{The dependence of the maximum resonant differential cross section (\ref{27}), (\ref{28}), (\ref{46}) (in units of $r_e^2$) on the square of the positron outgoing angle (Channel A) or the square of the electron outgoing angle (Channel B) for various intensities of the X-ray wave and the
characteristic energy of BWP, as well as the numbers of absorbed ($r$) and emitted ($r'$) photons in conditions (\ref{9}) for the energies of the initial gamma quanta. The energies of the initial gamma quanta:  Fig.3a) -- $\omega_2\gg\omega_{BW}=17.4\,\mbox{GeV},\,\omega_1=1\,\mbox{GeV}$; Fig.3b) and 3c) -- $\omega_2\gg\omega_{BW}=522\,\mbox{MeV},\,\omega_1=50\,\mbox{MeV}$.}
\label{fig4}
\end{figure}

\begin{figure}[H]
\begin{minipage}[h]{0.325\linewidth}
\center{\includegraphics[width=0.99\linewidth]{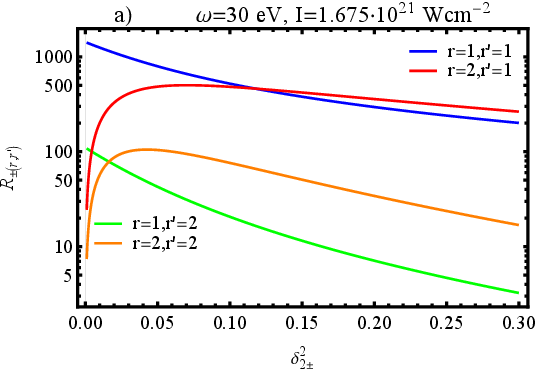}}
\end{minipage}
\hfill
\begin{minipage}[h]{0.325\linewidth}
\center{\includegraphics[width=0.99\linewidth]{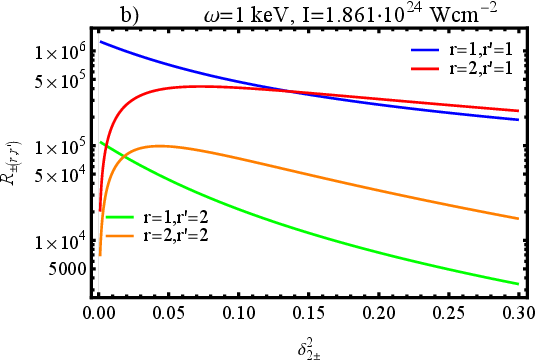}}
\end{minipage}
\hfill
\begin{minipage}[h]{0.325\linewidth}
\center{\includegraphics[width=0.99\linewidth]{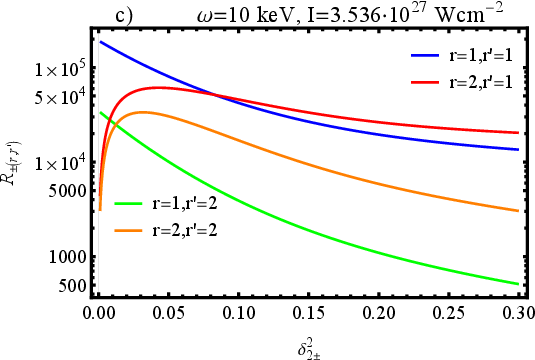}}
\end{minipage}
\caption{The dependence of the maximum resonant differential cross section (\ref{27}), (\ref{28}), (\ref{46}) (in units of $r_e^2$) on the square of the positron outgoing angle (Channel A) or the square of the electron outgoing angle (Channel B) for various intensities of the X-ray wave and the
characteristic energy of BWP, as well as the numbers of absorbed ($r$) and emitted ($r'$) photons in conditions (\ref{10}) for the energies of the initial gamma quanta. The energies of the initial gamma quanta:  Fig.3a) -- $\omega_2\gg\omega_{BW}=17.4\,\mbox{GeV},\,\omega_1=0.35\,\mbox{GeV}$; Fig.3b) and 3c) -- $\omega_2\gg\omega_{BW}=522\,\mbox{MeV},\,\omega_1=12\,\mbox{MeV}$.}
\label{fig5}
\end{figure}

 From equations (\ref{27}), (\ref{28}), and (\ref{33}), it can be seen that the magnitude of the resonant differential cross section is mainly determined by function $c_{\eta i}$ (\ref{28}). With a fixed value of the initial ultrarelativistic parameter $\delta^2_{\eta i}$ (\ref{29}), the function $c_{\eta i}$ is significantly dependent on the characteristic energy of BWP $\omega_{BW}$ (\ref{7}), as well as on function $K(\varepsilon_{1C})$ (\ref{30}) that determines the resonance width \cite{52}. As the characteristic energy of BWP decreases, the resonant cross section increases quite rapidly (see Fig.\ref{fig3} - Fig.\ref{fig5}).
\par Within the scope of the case of very high energies of the second gamma quantum (\ref{9}) studied in this article, the resonant energy of the positron (electron) takes the form (\ref{20}). Taking this into account, expressions (\ref{36})-(\ref{40}) will take the following form:
\begin{equation}\label{41}
\gamma_{2\pm}\approx4r\frac{\eta}{\sqrt{1+\eta^2}}\frac{\delta_{2\pm}}{1+4\delta_{2\pm}^2},
\end{equation}
\begin{equation}\label{42}
u_{2\pm}\approx\frac{r\varepsilon_{2BW}}{1+4\delta_{2\pm}^2}\gg1,
\end{equation}
\begin{equation}\label{43}
\gamma_{1\mp}\approx2r\frac{\eta}{\sqrt{1+\eta^2}}\frac{\sqrt{(r'/r)\beta_{\pm}-1}}{\beta_{\pm}},
\end{equation}
\begin{equation}\label{44}
\beta_{\pm}=(1+4\delta_{2\pm}^2)\left[1+\frac{1+4\delta_{2\pm}^2}{r\varepsilon_{1C}}\right],
\end{equation}
\begin{equation}\label{45}
u_{1\mp}\approx\left[1+\frac{1+4\delta_{2\pm}^2}{r\varepsilon_{1C}}\right]^{-1}.
\end{equation}
Due to this, the functions $\Psi_{\pm(rr')}$ (\ref{33}) in the expression for the resonant differential cross section (\ref{27}) is simplified and takes the following form:
\begin{equation}\label{46}
\Psi_{\pm(rr')}=\frac{8\eta^2r^2K_{1\mp(r')}}{(1+4\delta_{2\pm}^2+r\varepsilon_{1C})(1+4\delta_{2\pm}^2)}\left[\left(\frac{r^2}{\gamma_{2\pm}^2}-1\right)J^2_{r}(\gamma_{2\pm})+J'^2_{r}\right].
\end{equation}
Here, the argument of the Bessel functions $\gamma_{2\pm}$ takes the form (\ref{41}). It should be noted that the obtained resonant cross section (\ref{27}), (\ref{28}), (\ref{46}) depends on the frequency of the high-energy second gamma quantum (\ref{9}) only through the ultrarelativistic parameters (\ref{29}) and (\ref{18}).
\par Let us consider the case when the energies of the initial gamma quanta satisfy the conditions (\ref{10}). Then the parameter $u_{1\mp}$ (\ref{45}) takes the following form:
\begin{equation}\label{47}
u_{1\mp}\approx\frac{r\varepsilon_{1C}}{1+4\delta_{2\pm}^2}\ll1.
\end{equation}
As a result, expressions (\ref{35}) and (\ref{43}) are simplified:
\begin{equation}\label{48}
    K_{1\mp(r')}=-4J^2_{r'}(\gamma_{1\mp})+2\eta^2(J^2_{r'-1}+J^2_{r'+1}-2J^2_{r'}).
\end{equation}
\begin{equation}\label{49}
\gamma_{1\mp}\approx2r\frac{\eta}{\sqrt{1+\eta^2}}\frac{\sqrt{r'\varepsilon_{1C}}}{1+4\delta_{2\pm}^2}.
\end{equation}
Thus, the resonant cross section for the initial gamma quanta energies (\ref{10}) is given by equations (\ref{27}), (\ref{28}), (\ref{46}) in which the functions $K_{1\mp(r')}$, $\gamma_{1\mp}$ and $\gamma_{2\pm}$ are determined by relationships (\ref{48}), (\ref{49}), and (\ref{41}) respectively.
\par Figures \ref{fig3}, \ref{fig4}, and \ref{fig5} show the distributions of the resonant cross section in the strong X-ray field with varying numbers of absorbed and emitted photons as a function of the square of the positron (for Channel A) or electron (for Channel B) outgoing angle relative to the momentum of the second gamma quantum under the conditions (\ref{8}), (\ref{9}), and (\ref{10}) for various characteristic energies of BWP. The initial ultrarelativistic parameter $\delta^2_{\eta i}=10^{-2}$. Tables \ref{tab1}, \ref{tab2}, and \ref{tab3} represent the maximum values of the resonant cross sections $R^{*}_{\pm(r,r')}$ in the corresponding peaks of the distributions as functions of $\delta^2_{2\pm}$. From the data in the figures and corresponding tables, it is evident how the resonant cross sections change when the characteristic energy of BWP varies from $\omega_{BW}=17.4$\,GeV to $\omega_{BW}=522$\,MeV. In this case, the magnitude of $\omega_{BW}$ changes both due to the variation of the wave frequency and the parameter $\eta$ (see equation (\ref{11})). Thus, if the wave intensity is confined to the interval $I=1.675\cdot10^{21}\div1.861\cdot10^{24}\,(3.536\cdot10^{27})$ Wcm$^{-2}$, the maximum resonant cross sections (in units $r_e^2$) for $r=r'=1$ change as follows: under the conditions of the initial gamma quanta energies (\ref{8}) in the interval $R^{*}_{\pm(1,1)}\approx1.13\cdot10^{2}\div4.36\cdot10^{4}\,(1.41\cdot10^{4})$; under the conditions of the initial gamma quanta energies (\ref{9}) in the interval $R^{*}_{\pm(1,1)}\approx1.68\cdot10^{2}\div6.65\cdot10^{4}\,(2.00\cdot10^{4})$; under the  conditions of the initial gamma quanta energies (\ref{10}) in the interval $R^{*}_{\pm(1,1)}\approx1.43\cdot10^{3}\div1.27\cdot10^{6}\,(3.39\cdot10^{5})$. Therefore, when the characteristic energy of BWP is reduced by a factor of 33, the maximum resonant cross section increases by a factor of $3.86\cdot10^{2}(1.25\cdot10^{2})$ under the conditions of the initial gamma quanta energies (\ref{8}), by a factor of $3.96\cdot10^{2}(1.19\cdot10^{2})$ under the conditions of the initial gamma quanta energies (\ref{9}), and by a factor of $0.89\cdot10^{3}(2.37\cdot10^{2})$ under the conditions of the initial gamma quanta energies (\ref{10}).
\par Consequently, the maximum value of the resonant cross section occurs under the conditions of the initial gamma quanta energies (\ref{10}) $(\omega_1\ll\omega_C,\,\omega_2\gg\omega_{BW})$ and, depending on the value of the characteristic energy of BWP in the interval $\omega_{BW}\sim10\div1$ GeV, can reach a magnitude of $R^{*}_{\pm(1,1)}\sim10^{3}\div10^{6}\,r_e^2$. At the same time, for Channel A, narrow beams of high-energy positrons $(E_+\rightarrow\omega_2)$ are obtained, while for Channel B, narrow beams of high-energy electrons $(E_-\rightarrow\omega_2)$ are obtained (see relation (\ref{20})).
\par However, if the characteristic BWP energy remains unchanged with changes in the frequency and intensity of the wave, then the resonant cross section changes less significantly (compare the values of the resonant cross section without parentheses and within parentheses in the above text).

\begin{table}[H]
\centering
\begin{tabular}{|c|c|c|c|}
\hline
                  & (r,r') & $\delta^{2(*)}_{2\pm}$ & $R^{(*)}_{\pm(r,r')}$ \\ \hline
\multirow{4}{*}{\begin{tabular}{c}
$I=1.675\cdot10^{21}\,\mbox{Wcm}^{-2}$,\\
$\omega_1=1\,\mbox{GeV}$,\\
$\omega_2=35\,\mbox{GeV}$\end{tabular}} & (1,1)       & 0  & $1.13\cdot10^2$  \\ \cline{2-4} 
                 &   (2,1)     &  0.095 &  $3.90\cdot10^1$ \\ \cline{2-4} 
                  &   (1,2)     &  0 & $1.69\cdot10^1$  \\ \cline{2-4} 
                  &    (2,2)    & 0.060  & $1.53\cdot10^1$  \\ \hline
\multirow{4}{*}{\begin{tabular}{c}
$I=1.861\cdot10^{24}\,\mbox{Wcm}^{-2}$,\\
$\omega_1=50\,\mbox{MeV}$,\\
$\omega_2=1\,\mbox{GeV}$\end{tabular}} &(1,1)        &  0 & $4.36\cdot10^4$  \\ \cline{2-4} 
                  &   (2,1)     &  0.047 & $1.66\cdot10^4$  \\ \cline{2-4} 
                  &   (1,2)     &  0 & $9.85\cdot10^3$  \\ \cline{2-4} 
                  &   (2,2)     & 0.069  & $6.83\cdot10^3$  \\ \hline
\multirow{4}{*}{\begin{tabular}{c}
$I=3.536\cdot10^{27}\,\mbox{Wcm}^{-2}$,\\
$\omega_1=50\,\mbox{MeV}$,\\
$\omega_2=1\,\mbox{GeV}$\end{tabular}} & (1,1)       & 0  &  $1.41\cdot10^4$ \\ \cline{2-4} 
                  &   (2,1)     & 0.025  & $6.41\cdot10^3$  \\ \cline{2-4} 
                  &    (1,2)    & 0  &$6.13\cdot10^3$   \\ \cline{2-4} 
                  &   (2,2)     & 0.048  &  $3.49\cdot10^3$ \\ \hline
\end{tabular}
\caption{The values of the ultrarelativistic parameter $\delta^{2(*)}_{2\pm}$ that correspond maximum 
values $R^{(*)}_{\pm(r,r')}$ of the spectral-angular distribution for the resonant differential cross sections (\ref{27}) under the conditions (\ref{8}) (see Fig.\ref{fig3} and expression (\ref{11}))}
\label{tab1}
\end{table}

\begin{table}[H]
\centering
\begin{tabular}{|c|c|c|c|}
\hline
                  & (r,r') & $\delta^{2(*)}_{2\pm}$ & $R^{(*)}_{\pm(r,r')}$ \\ \hline
\multirow{4}{*}{\begin{tabular}{c}
$I=1.675\cdot10^{21}\,\mbox{Wcm}^{-2}$,\\
$\omega_1=1\,\mbox{GeV}$\end{tabular}} & (1,1)       & 0  & $1.68\cdot10^2$  \\ \cline{2-4} 
                 &   (2,1)     &  0.070 &  $4.66\cdot10^1$ \\ \cline{2-4} 
                  &   (1,2)     &  0 & $3.40\cdot10^1$  \\ \cline{2-4} 
                  &    (2,2)    & 0.055  & $1.99\cdot10^1$  \\ \hline
\multirow{4}{*}{\begin{tabular}{c}
$I=1.861\cdot10^{24}\,\mbox{Wcm}^{-2}$,\\
$\omega_1=50\,\mbox{MeV}$\end{tabular}} &(1,1)        &  0 & $6.65\cdot10^4$  \\ \cline{2-4} 
                  &   (2,1)     &  0.033 & $2.59\cdot10^4$  \\ \cline{2-4} 
                  &   (1,2)     &  0 & $2.02\cdot10^4$  \\ \cline{2-4} 
                  &   (2,2)     & 0.063  & $8.84\cdot10^3$  \\ \hline
\multirow{4}{*}{\begin{tabular}{c}
$I=3.536\cdot10^{27}\,\mbox{Wcm}^{-2}$,\\
$\omega_1=50\,\mbox{MeV}$\end{tabular}} & (1,1)       & 0  &  $2.00\cdot10^4$ \\ \cline{2-4} 
                  &   (2,1)     & 0.020  & $1.15\cdot10^4$  \\ \cline{2-4} 
                  &    (1,2)    & 0  &$1.18\cdot10^4$   \\ \cline{2-4} 
                  &   (2,2)     & 0.043  &  $4.19\cdot10^3$ \\ \hline
\end{tabular}
\caption{The values of the ultrarelativistic parameter $\delta^{2(*)}_{2\pm}$ that correspond maximum 
values $R^{(*)}_{\pm(r,r')}$ of the spectral-angular distribution for the resonant differential cross sections (\ref{27}) under the conditions (\ref{9}) (see Fig.\ref{fig4} and expression (\ref{11}))}
\label{tab2}
\end{table}

\begin{table}[H]
\centering
\begin{tabular}{|c|c|c|c|}
\hline
                  & (r,r') & $\delta^{2(*)}_{2\pm}$ & $R^{(*)}_{\pm(r,r')}$ \\ \hline
\multirow{4}{*}{\begin{tabular}{c}
$I=1.675\cdot10^{21}\,\mbox{Wcm}^{-2}$,\\
$\omega_1=0.35\,\mbox{GeV}$\end{tabular}} & (1,1)       & 0  & $1.43\cdot10^3$  \\ \cline{2-4} 
                 &   (2,1)     &  0.070 &  $5.01\cdot10^2$ \\ \cline{2-4} 
                  &   (1,2)     &  0 & $1.10\cdot10^2$  \\ \cline{2-4} 
                  &    (2,2)    & 0.043  & $1.05\cdot10^2$  \\ \hline
\multirow{4}{*}{\begin{tabular}{c}
$I=1.861\cdot10^{24}\,\mbox{Wcm}^{-2}$,\\
$\omega_1=12\,\mbox{MeV}$\end{tabular}} &(1,1)        &  0 & $1.27\cdot10^6$  \\ \cline{2-4} 
                  &   (2,1)     &  0.073 & $2.33\cdot10^5$  \\ \cline{2-4} 
                  &   (1,2)     &  0 & $1.12\cdot10^5$  \\ \cline{2-4} 
                  &   (2,2)     & 0.044  & $9.88\cdot10^4$  \\ \hline
\multirow{4}{*}{\begin{tabular}{c}
$I=3.536\cdot10^{27}\,\mbox{Wcm}^{-2}$,\\
$\omega_1=12\,\mbox{MeV}$\end{tabular}} & (1,1)       & 0  &  $3.39\cdot10^5$ \\ \cline{2-4} 
                  &   (2,1)     & 0.073  & $1.08\cdot10^5$  \\ \cline{2-4} 
                  &    (1,2)    & 0  &$6.11\cdot10^4$   \\ \cline{2-4} 
                  &   (2,2)     & 0.044  &  $5.92\cdot10^4$ \\ \hline
\end{tabular}
\caption{The values of the ultrarelativistic parameter $\delta^{2(*)}_{2\pm}$ that correspond maximum 
values $R^{(*)}_{\pm(r,r')}$ of the spectral-angular distribution for the resonant differential cross sections (\ref{27}) under the conditions (\ref{10}) (see Fig.\ref{fig5} and expression (\ref{11}))}
\label{tab3}
\end{table}

\section*{Conclusion}
\par We considered the resonant BWP modified by a strong X-ray field for high-energy initial gamma quanta. The following results were obtained:
\begin{enumerate}
    \item It is shown that the resonant cross section significantly depends on the magnitude of the characteristic Breit-Wheeler energy $\omega_{BW}$ (\ref{7}) and the characteristic energy of the Compton effect $\omega_C$ (\ref{5}). The ratios of the initial energies of gamma quanta with these characteristic energies significantly affect the magnitude of the resonant cross section.
    \item Under the conditions when the energy of the second gamma quantum significantly exceeds the characteristic Breit-Wheeler energy $(\omega_2\gg\omega_{BW})$, the resonant energy of the positron (for Channel A) or electron (for Channel B) tends to the energy of the high-energy second gamma quantum (\ref{20}) $(E_{\pm}\rightarrow\omega_2)$.
    \item The magnitude of the resonant cross section significantly depends on the characteristic Breit-Wheeler energy as well as the width of the resonance $(R_{\pm(rr')}\sim\omega^{-2}_{BW}K^{-2}(\varepsilon_{1C}))$ (see equations (\ref{27}) and (\ref{28})). Consequently, by decreasing the characteristic Breit-Wheeler energy by one order of magnitude, the resonant cross section increases by two orders of magnitude. Moreover, the highest resonant differential cross section is achieved when the energy of the second gamma quantum significantly exceeds the characteristic Breit-Wheeler energy $(\omega_2\gg\omega_{BW})$, while the energy of the first gamma quantum is significantly lower than the characteristic Compton effect energy $(\omega_1\ll\omega_C)$. In this case, when reducing the characteristic BWP energy within the range $\omega_{BW}\sim10\div1$ GeV, the maximum resonant cross section can reach values of $R^{*}_{\pm(1,1)}\sim10^{3}\div10^{6}\,r_e^2$.
\end{enumerate}
\par The obtained results could be used to explain the narrow fluxes of high-energy ultrarelativistic positrons (electrons) in the vicinity of neutron stars and magnetars, as well as in modeling physical processes in laser thermonuclear fusion.
\par The research was funded by the Ministry of Science and Higher Education of the Russian Federation under the strategic academic leadership program “Priority 2030” (Agreement 075-15-2023-380 dated 20.02.2023).

\end{document}